\documentclass{emulateapj}
\usepackage{apjfonts}

\shorttitle{Accurate Position for HDF~850.1}

\begin{document}

\title{An Accurate Position for HDF~850.1:
The Brightest Submillimeter Source in the Hubble Deep Field-North}

\author{L. L. Cowie\altaffilmark{1}, 
A. J. Barger\altaffilmark{2,3,1},
W.-H. Wang\altaffilmark{4,5,6}, 
J. P. Williams\altaffilmark{1}}

\altaffiltext{1}{Institute for Astronomy, University of Hawaii,
2680 Woodlawn Drive, Honolulu, HI 96822.}
\altaffiltext{2}{Department of Astronomy, University of Wisconsin-Madison,
475 N. Charter Street, Madison, WI 53706.}
\altaffiltext{3}{Department of Physics and Astronomy, University of Hawaii,
2505 Correa Road, Honolulu, HI 96822.}
\altaffiltext{4}{Jansky Fellow.}
\altaffiltext{5}{National Radio Astronomy Observatory,
1003 Lopezville Road, Socorro, NM 87801. The NRAO is a facility 
of the National Science Foundation operated under cooperative 
agreement by Associated Universities, Inc.}
\altaffiltext{6}{Current address: Institute of Astronomy and Astrophysics, 
Academia Sinica, PO Box 23-141, Taipei 10617, Taiwan.}

\slugcomment{Astrophysical Journal Letters 697 (2009) L122-L126}

\begin{abstract}

We report a highly significant Submillimeter Array (SMA) detection 
of the prototypical submillimeter source HDF~850.1, which is the 
brightest submillimeter source in the Hubble Deep Field-North 
proper. The detection yields an extremely precise position of 
RA(2000)=12$^h$~36$^m$~51.99$^s$
and Dec(2000)=+62$^\circ$~12$^{\prime}$~25.83$^{\prime \prime}$ 
with a $1\sigma$ positional uncertainty of $0\farcs17$. 
The position is consistent with the location of a millimeter 
wavelength interferometric detection and with the locations of weak 
Very Large Array detections at 1.4 and 8.4~GHz, but it is not 
consistent with any previous optical/near-infrared identifications. 
The source appears pointlike at the 
$2^{\prime \prime}$ resolution of the SMA, and the detected flux 
of $7.8\pm1.0$~mJy is consistent with the measured 
Submillimeter Common-User Bolometer Array (SCUBA) fluxes.
We tabulate fluxes and limits on HDF~850.1 at other wavelengths.
Our redshift estimate for HDF~850.1 based on the radio 
through mid-infrared measurements is $z=4.1^{+0.5}_{-0.6}$. 
The faintness of the source at optical/near-infrared wavelengths 
and the high estimated redshift suggest that 
HDF~850.1 may be an analog of the brighter submillimeter source 
GOODS~850-5, which is also thought to be at $z>4$. 
The fact that a source like HDF~850.1
should have appeared in one of the very first blank-field SCUBA 
observations ever made suggests that such high-redshift sources 
are quite common. Thus, we are led to conclude that 
high-redshift star formation is dominated by giant dusty
star-forming galaxies, just as it is at lower redshifts.

\end{abstract}


\section{Introduction}

In the more than ten years since the first discoveries 
(Smail et al.\ 1997; Barger et al.\ 1998; Hughes et al.\ 1998; Eales et al.\ 1999) 
of distant submillimeter galaxies were made using the Submillimeter
Common-User Bolometer Array (SCUBA; Holland et al.\ 1999) 
on the single-dish James Clerk Maxwell Telescope, 
we have learned a great deal about these sources and their contribution
to the 850~$\mu$m background light. We now know the 
background is dominated by sources with 850~$\mu$m fluxes near 1~mJy 
(Blain et al.\ 1999; Cowie et al.\ 2002; Knudsen et al.\ 2008).
In addition, stacking analyses suggest that much of this light arises
from galaxies near redshift one (Wang et al.\ 2006; Serjeant et al.\ 2008). 
The large positional uncertainties of the SCUBA sources 
make direct spectroscopic measurements of potential
optical counterparts time-consuming 
and ambiguous (Barger et al.\ 1999). However, many 
($\sim60-70$\%; Barger et al.\ 2000; Ivison et al.\ 2002;
Chapman et al.\ 2003b)
of the SCUBA sources with 850~$\mu$m fluxes
above 5~mJy have 1.4~GHz counterparts
whose positions are known with subarcsecond accuracy.
The optical and near-infrared (NIR) counterparts to 
many of these have been spectroscopically identified, most
of which are found to lie in the redshift range $z=2-3$ 
(Chapman et al.\ 2003a, 2005). 

Throughout the redshift range $z=1-3$ the submillimeter sources 
dominate the universal star formation history
(e.g., Chapman et al.\ 2005; Wang et al.\ 2006).
The key questions are whether this continues to higher redshifts,
and how high in redshift one must go before one finds that smaller 
galaxies are dominating the universal star formation history. 
If giant submillimeter galaxies
still make up the bulk of the star formation at very high redshifts
($z\sim5$ or more), this could pose a severe challenge to
the cold dark matter models of galaxy growth. At present there are 
only a very small number of spectroscopically identified submillimeter
galaxies at $z>3$,
with the current highest redshifts being at $z\sim4.5-4.7$ 
(Capak et al.\ 2008; Coppin et al.\ 2009).
However, this may in large part be an observational selection effect.
First, unlike the submillimeter with its negative $K$-correction, the 
radio dims at higher redshifts.
Even at $>5$~mJy roughly $30-40$\% of the sources
do not have strong radio counterparts. 
Since these sources cannot be spatially localized, there is no 
simple route to obtaining a redshift.
Second, it is the high-redshift sources that are most likely to be 
omitted from the spectroscopic samples, since they are likely to
be optically fainter.
Thus, while it is clear that submillimeter galaxies could
continue to dominate the star formation history at $z>3$,
we are still in the uncomfortable position of not knowing the 
exact positions or counterparts of most of the submillimeter sources
that are likely to lie at these redshifts. 

Fortunately, the advent of the Submillimeter Array 
(SMA; Ho et al.\ 2004) has allowed
accurate positions to be obtained for some of these sources
(Iono et al.\ 2006; Wang et al.\ 2007; Younger et al.\ 2007).
Of particular interest is the source GOODS~850-5, which
may lie at a very high redshift (e.g., Wang et al.\ 2007, 2009;
Dannerbauer et al.\ 2008).
Daddi et al.\ (2009a) place GOODS~850-5 at $z=4.1$ based
on a possible identification of a single CO line.

All of the SMA-detected sources are relatively bright, but it 
is now possible, with the improving sensitivity of the SMA, 
to study more typical sources. In the present paper we report 
on SMA observations of HDF~850.1 (Hughes et al.\ 1998). 
HDF~850.1 is the strongest submillimeter source in the Hubble
Deep Field-North (HDF-N) proper,
and, as such, it can almost be considered as the prototypical SCUBA
source. It was also among the first SCUBA-selected sources to be detected 
with millimeter wavelength interferometry,
giving an accurate position with an uncertainty of about  
$0\farcs3$ (Downes et al.\ 1999).
It is a measure of the difficulty of identifying the SCUBA sources
that, despite this measurement and 
the extremely deep optical, NIR, and radio data available for the
HDF-N, no convincing optical/NIR counterpart has been found for HDF~850.1. 
Indeed, several possible optical/NIR counterparts have been suggested, with
the most recent identification being
put forward by Dunlop et al.\ (2004). 

We shall show in the present paper that none of the previous 
optical/NIR identifications is correct and that HDF~850.1 has no detectable 
optical or NIR light. It is a weak radio source with detections 
at both 1.4 and 8.4~GHz, and it may also be detected at 24~$\mu$m. 
Overall, the far-infrared (FIR) and radio spectral energy 
distributions (SEDs) would place HDF~850.1 at a redshift of just 
above $z=4$, if its rest-frame SED is similar to that of 
low-redshift ultraluminous infrared galaxies (ULIRGs).

Throughout this paper we assume 
$H_0 = 70$~km~s$^{-1}~{\rm Mpc^{-1}}$, $\Omega_M = 0.3$, 
and $\Omega_{\Lambda} = 0.7$.

%
%
%
\begin{figure}
\epsscale{1.0}
\plotone{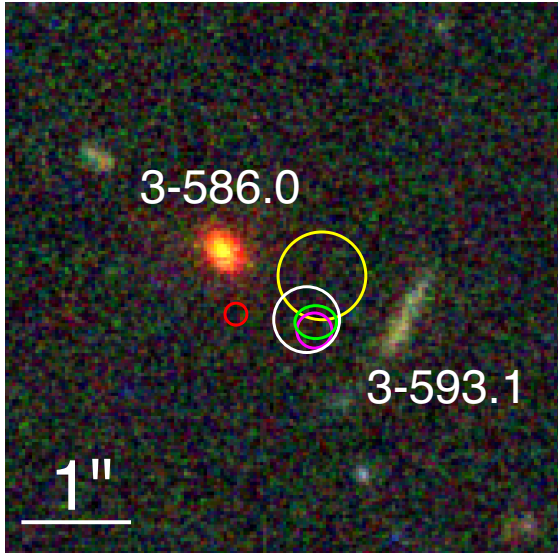}
\caption{
The measured positions for HDF~850.1 are shown
superimposed on an Advanced Camera for Surveys three-color 
image (B: F435W; G: F606W; R: F814W$+$F850LP; 
Giavalisco et al.\ 2004). The error ellipses for the SMA (green),
IRAM (white), VLA 1.4~GHz (purple),
and VLA 8.4~GHz (yellow) observations are shown. 
The red circle shows the position of the counterpart suggested 
by Dunlop et al.\ (2004), which is significantly rejected by the
current observations. The two neighboring galaxies are
3-586.0 (the red elliptical galaxy) and 3-593.1 
(the blue arc-like galaxy) in the catalog of Williams et al.\ (1996).
}
\label{fig1}
\end{figure}

%
%
\begin{figure}
\epsscale{1.0}
\plotone{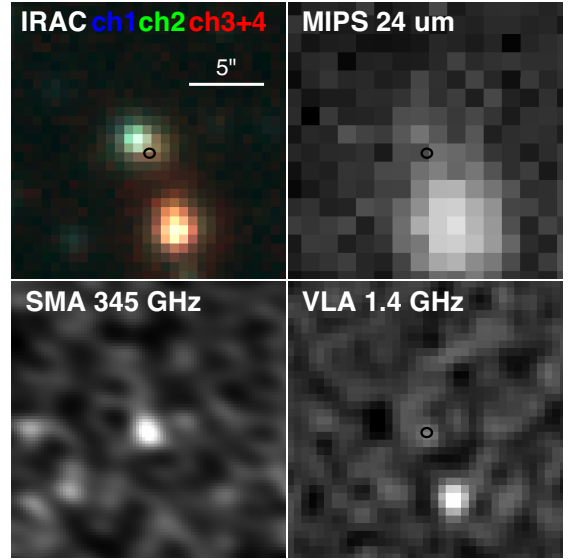}
\caption{Mid-infrared to radio images of HDF~850.1 centered
on the SMA position. The black small circles show the SMA 
position and its uncertainty.
The IRAC and MIPS images are from the
Great Observatories Origins Deep Surveys \emph{Spitzer} 
Legacy Program (M.~Dickinson et al.\ 2009, in preparation).
The radio image is provided by G.~Morrison (2009, private 
communication). 
}
\label{fig2}
\end{figure}

\section{SMA Observations}

Two full tracks of SMA observations of HDF~850.1 were obtained in 
2008 February and April with, respectively, eight antennas in the 
compact configuration and seven antennas in the compact-north 
configuration.  The receivers were tuned such that the central 
frequency of the observations was at 345~GHz.  Callisto and Ceres 
were used as flux calibrators, and 3C454.3 was used as the bandpass 
calibrator.  Quasars 1419+543 and 1048+717, which are, respectively,
$15.\mkern-6mu\arcdeg5$ and $14\arcdeg$ away from the target, 
were observed after every 15 minutes of on-target integration for 
time-dependent complex gain calibrations.  The averaged single-sideband 
system temperatures in the two tracks were 420 and 620~K.

The calibration and data inspection were performed with the Caltech 
package MIR modified for the SMA.  Continuum data were generated
by averaging the spectral channels after the passband calibration.
Both gain calibrators were used to derive gain curves. 
Flux calibrations were performed using data taken under conditions 
(time, hour angle, and elevation) similar to that of the flux calibrator.
The error in flux calibration is usually within 10\% with this method.
The calibrated visibility data were Fourier transformed and
deconvolved in the package MIRIAD to form images.  In the 
transformation we applied the ``robust weighting'' of Briggs (1995),
with a robust parameter of +0.9 to obtain a better balance 
between beam size and S/N.  We also weighted each visibility 
point inversely proportional to the system temperature.  
The synthesized beam has a FWHM of $2\farcs09 \times 1\farcs76$ 
at a position angle of $60\arcdeg$.  The theoretical noise and the 
noise measured in the CLEAN deconvolved image are 0.94 and 
1.05~mJy, respectively.

HDF~850.1 is detected in both the dirty and CLEANed images and is
unresolved. In MIRIAD a point-source fit to the image yields a flux of
7.8~mJy and a J2000 position of \\

\noindent
RA 12$^h$ 36$^m$ 51.99$^s$ ($\pm0\farcs18$)\\
Dec +62$^\circ$ 12$^{\prime}$ 25.83$^{\prime \prime}$ ($\pm0\farcs16$) \,. 
\\

\noindent
Here the positional errors were obtained using the imfit routine
in MIRIAD with a box size equal to 2.5 times the beam size. A more
direct method of finding the position and error is to fit the
visibilities. Fitting with the uvfit routine in MIRIAD gives
a flux of $8.2\pm1.6$~mJy and a position of 

\noindent
RA 12$^h$ 36$^m$ 51.97$^s$ ($\pm0\farcs19$)\\
Dec +62$^\circ$ 12$^{\prime}$ 25.69$^{\prime \prime}$ ($\pm0\farcs15$) \,.
\\

\noindent
The weighting of the flux extraction with uvfit may be less optimal
than that with the imfit routine, so we adopt the imfit estimate
of the flux and error. However, the error estimate in the imfit
position is sensitive to the choice of box size, while the uvfit
errors are not sensitive to such assumptions. We therefore adopt
the uvfit position and positional errors.

In the images of the gain calibrators the measured positional offsets
are all within the above errors.  Thus, we do not consider the
calibration error to be a significant source of error in the astrometry.
More specifically, we note that the observations of the two quasars
were interleaved with each other and that we used the measured phase
of each to derive the astrometry. They are not each self-calibrated,
i.e., placed at their VLA position. Combining the phases for both
gives more accurate positions over a wider region of
sky. We can make an estimate of the systematics 
by measuring the positional offsets
of each quasar relative to their nominal positions, obtaining
($0\farcs17$,$-0\farcs06$) for 1419+545 and 
($-0\farcs03$,$-0\farcs06$) for 1048+717.
Since HDF850.1 lies between the two quasars, the positional
offset will, at some level, average these values. However,
if we use the maximum values here as our estimate of
the systematic error, our positional error becomes $0\farcs25$
in RA and $0\farcs16$ in Dec. We will use this error estimate,
which we believe to be quite conservative, in the subsequent discussion.

The SMA flux of HDF~850.1 is consistent with the earliest SCUBA
jiggle-map measurement of Hughes et al.\ (1998), who found a flux
density of $S_{\rm 850~\mu m} = 7.0 \pm 0.5$~mJy, but it is larger than 
all of the later SCUBA results obtained with various mapping and
source extraction methods (Serjeant et al.\ 2003; Borys et al.\ 2003;
Wang et al.\ 2004). It lies about $2''$ from the SCUBA position of 
Serjeant et al.\ (2003), which is reasonable given the expected positional 
error for a SCUBA source with this S/N (e.g., Wang et al.\ 2004).

\section{Other Measurements of HDF~850.1}

\subsection{Millimeter and Radio}

In Table~\ref{tab1} we list the other millimeter and radio measurements
(plus references) of HDF 850.1.
In Figure~\ref{fig1} we show the relative locations of some of those 
measurements with circles indicating the 1$\sigma$ positional 
uncertainties. The SMA position is shown with the green ellipse. It is 
fully consistent with the 1.35~mm measurement of 
Downes et al.\ (1999) obtained using the IRAM PdB with a
J2000 position (white circle in Fig.~\ref{fig1}) of \\ 

\noindent
RA 12$^h$~36$^m$~51.98$^s$ \\
Dec +62$^\circ$~12$^{\prime}$~25.7$^{\prime \prime}$ \,.
\\

\noindent
The beam for this measurement was approximately $2''$, and the 
published $1\sigma$ positional uncertainty of $0\farcs3$ seems
very  conservative. In this regard the IRAM center lies only
$0\farcs07$ from our SMA position, well within the quoted errors.
(If we use the image plane fit as opposed to the UV plane fit,
this offset rises to $0\farcs19$, which is also well within the errors.)

The  source is also weakly detected at 1.4 and 8.4~GHz. There is
a nearly $4\sigma$ detection at 1.4~GHz in the deep map of the
area obtained by G.~Morrison et al.\ (2009, in preparation). The flux
is $16.73\pm4.25$~$\mu$Jy and the J2000 position 
(purple circle in Fig.~\ref{fig1}) is \\

\noindent
RA 12$^h$ 36$^m$ 51.97$^s$ \\ 
Dec +62$^\circ$ 12$^{\prime}$ 25.6$^{\prime \prime}$ \,, \\

\noindent
with a positional uncertainty of $0\farcs16$
(G.~Morrison, private communication).
This source is also detected at the 3.5$\sigma$ level at 
8.4~GHz (VLA~3651+1226 in the supplementary 
list of sources in Richards et al.\ 1998). 
The 8.4~GHz J2000 position (yellow circle in Fig.~\ref{fig1}) is \\ 

\noindent
RA(2000) 12$^h$ 36$^m$ 51.96$^s$ \\ 
Dec(2000) +62$^\circ$ 12$^{\prime}$ 26.1$^{\prime \prime}$ \,, \\

\noindent
which is consistent with the SMA position within the fairly substantial
($\sim 0\farcs4$) positional uncertainty for this low significance source.
We shall use the 8.4~GHz flux of $7.5\pm2.2$~$\mu$Jy quoted by 
Dunlop et al.\ (2004) as a private communication from E. Richards.
Given the low signal to noise of both the radio detections, we do
not use their positional information in any of the subsequent
discussion.

\subsection{Mid-infrared}

HDF~850.1 is not detected in \emph{Spitzer} MIPS 70~$\mu$m 
imaging (Huynh et al.\ 2007), so we adopt the nominal 2~mJy 
(3$\sigma$) upper limit. At 24~$\mu$m (see Fig.~\ref{fig2}) its 
flux is contaminated by a nearby radio source and possibly by the
elliptical galaxy 3-856.0 from the Williams et al.\ (1996) catalog.  
We used the \emph{Spitzer} 24~$\mu$m 
in-flight point-spread function (PSF) to subtract the flux from 
the nearby radio source and then measured a $7\arcsec$ 
aperture-corrected 
flux at the SMA position of HDF~850.1. The result is 
$28.2\pm4.4$~$\mu$Jy. This is consistent with the value in 
Pope et al.\ (2006). This may be considered as an upper limit,
since the contamination from 3-856.0 is unclear and the 
resolution of \emph{Spitzer} does not allow a reliable 
separation between the two. We note that the level of contamination 
from 3-856.0 in the IRAC bands (see Fig.~\ref{fig2}) is far 
too high to make any plausible flux estimates.

\subsection{Optical and Near-infrared}

As can be seen from Figure~\ref{fig1}, HDF~850.1 lies between 
two neighboring galaxies:  a red elliptical galaxy (3-586.0)
and a blue arc-like galaxy (3-593.1). 
The elliptical galaxy lies at a spectroscopic redshift of $z=1.224$
(Barger et al.\ 2008) and has no strong emission lines.
Dunlop et al.\ (2004) estimate a  velocity dispersion of
$\sigma_v = 146 \pm 29~{\rm km~s}^{-1}$ for this galaxy.
The blue arc-like galaxy is very faint and has not been 
spectroscopically observed.  Photometric redshift estimates 
place it at $z\simeq1.75$ (Fern{\'a}ndez-Soto et al.\ 1999;
Rowan-Robinson 2003). 
Both of these sources have been suggested as 
counterparts to HDF~850.1, but the present high-precision observations 
give separations that are clearly too large for this to be the case. 
Similarly, the object HDF~850.1K, which Dunlop et al.\ (2004) claim to 
be the counterpart (red circle in Fig.~\ref{fig1}), 
lies more than $0\farcs7$ from the SMA and IRAM source
positions. This identification is rejected at the $2.9\sigma$ 
level on positional grounds by the SMA data (using
the $0\farcs25$ RA error, since the separation is primarily along
this axis) and at the $2.2\sigma$ level by the IRAM
data. In combination, this identification is rejected at the $3.6\sigma$ level.
We conclude that HDF~850.1 has no detectable optical or NIR light.
We do not attempt to measure the optical and NIR fluxes at the SMA
position, as was done in Wang et al.\ (2009) at the SMA position
of GOODS~850-5, due to the inevitable contamination of any such 
measurements from the elliptical galaxy light.

%
%
\begin{deluxetable}{lcl}
\tablewidth{0pt}
\tablecaption{\small Mid-Infrared to Radio Fluxes for HDF~850.1}
\tablehead{Band & Flux & Reference}
\startdata
 $S_{\rm 24~\mu m}$ & $\phantom{<2}28.2 \pm 4.4$~$\mu$Jy & MIPS; this work\\
 $S_{\rm 70~\mu m}$ & $<2$ $(3\sigma)$~mJy & MIPS; Huynh et al.\ (2007)\\
 $S_{\rm 450~\mu m}$ & $<21$ $(3\sigma)$~mJy & SCUBA; Hughes et al.\ (1998)\\
 $S_{\rm 850~\mu m}$ & $\phantom{<2}7.0 \pm 0.4$~mJy & SCUBA; Hughes et al.\ (1998)\\
 $S_{\rm 1.3~mm}$ & $\phantom{<2}2.2 \pm 0.3$~mJy & IRAM; Downes et al.\ (1999)\\
 $S_{\rm 1.35~mm}$ & $\phantom{<2}2.1 \pm 0.5$~mJy & SCUBA; Hughes et al.\ (1998)\\ 
 $S_{\rm 8.4~GHz}$ & $\phantom{<2}7.5 \pm 2.2$~$\mu$Jy & E. Richards (2004, priv. comm.)\\
 $S_{\rm 1.4~GHz}$ & $\phantom{<2}16.73 \pm 4.25$~$\mu$Jy & G. Morrison (2009, priv. comm.)
\enddata
\label{tab1}
\end{deluxetable}

\section{SED and Redshift of HDF~850.1}

%
%
\begin{figure}
\epsscale{1.2}
\plotone{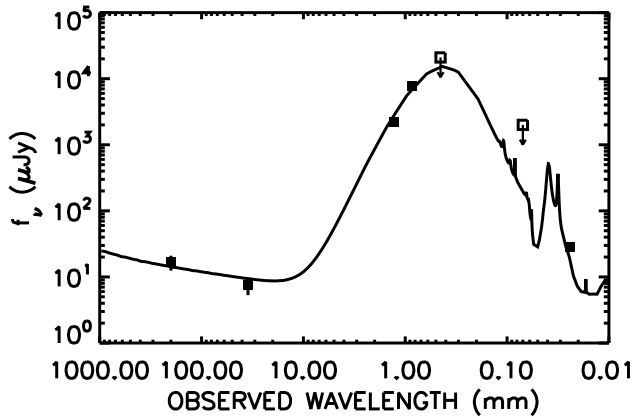}
\caption{
Radio to mid-infrared SED of HDF~850.1.  Solid squares 
are detections with 1$\sigma$ error bars.  Open squares 
with downward pointing arrows are 3$\sigma$ upper limits. 
The references for the data points are given in 
Table~\ref{tab1}. The curve shows the Arp~220 SED at
$z=4.1$. The SED is from Silva et al. (1998) but with
the radio fluxes fitted with a power law of the form
$f_{\nu} \sim \nu^{-0.38}$ rather than with the steeper slope
used in that paper.	
}
\label{fig3}
\end{figure}

In Figure~\ref{fig3} we show the observed SED of HDF~850.1 
at radio through mid-infrared wavelengths. We fitted a 
variety of models to determine the optimal template and 
redshift of the source. We show the Arp~220 template at 
$z=4.1$, which provides the best fit to the data,
as the solid curve in Figure~\ref{fig3}.
This gives the most likely redshift of 
HDF~850.1 as $z=4.1$ with a 68\% confidence interval of 
$3.5\le z \le 4.6$. This redshift estimate for HDF~850.1 is 
similar to the one made by Dunlop et al.\ (2004).  
However, the new 24~$\mu$m 
measurement (or upper limit) provides a much tighter constraint 
on possible templates, reducing the redshift range slightly.  
It is possible that HDF~850.1 lies in the redshift sheet
at $z\sim4.1$, which Daddi et al.\ (2009b) have found in the
GOODS-N region.

Wagg et al.\ (2007) did not detect any emission from HDF~850.1
in their Green Bank Telescope (GBT) wide bandwidth search for 
CO(1--0) from $z\sim3.3$--5.4 and for CO(2--1) from 
$z\sim3.9$--4.3. However, the sensitivity of the GBT is not
sufficient to place a strong limit on the CO line strengths 
relative to the infrared luminosity of HDF~850.1 based on the 
CO--FIR correlation of high-redshift ULIRGs (see 
Fig.~6 of Wagg et al.\ 2007). Thus, the estimated redshift 
of $z\sim4$ is not ruled out by the GBT non-detection.

We note in passing that Hughes et al.\ (1998) and Dunlop et al.\ (2004)
both suggested that HDF~850.1 may be being gravitationally 
lensed by the elliptical galaxy 3-586.0. While the effects of
gravitational lensing may brighten HDF~850.1 to some extent,
since we are not trying to use its flux to obtain a star formation
rate, we do not discuss the possible lensing effects here.

\section{Summary and Final Remarks}

%
%
%
\begin{figure}
\epsscale{1.2}
\plotone{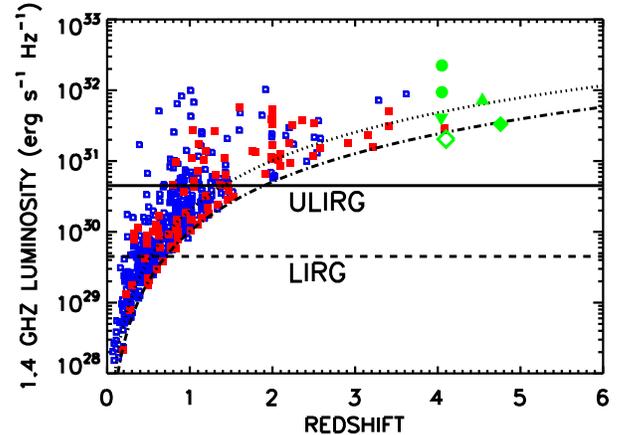}
\caption{
The radio powers of HDF~850.1 (green open diamond), 
GOODS~850-5 (green inverted triangle),
COSMOS J100054+023436 (green triangle),
GN20 and GN20.2 (green circles), and
LESS J033229.4-275619 (green solid diamond) vs. redshift
compared with the radio powers of spectroscopically
identified radio-selected sources in the GOODS-N, SSA13, and 
CLANS fields. Optically bright sources ($R<24.5$) are
denoted by blue open squares, and optically faint 
sources ($R>24.5$) by red solid squares.
The dashed and solid horizontal lines, respectively, show the 
radio powers corresponding to the FIR luminosities of luminous 
infrared galaxies (LIRGs;  $10^{11}~L_\odot\le L_{FIR}<10^{12}~L_\odot$)
and ULIRGs ($L_{FIR}\ge 10^{12}~L_\odot$), as determined by assuming 
the local FIR-radio correlation (see Barger et al.\ 2007 for our calculation).
The dot-dashed and dotted curves show the radio power limits 
corresponding to radio samples with 20~$\mu$Jy and 40~$\mu$Jy 
sensitivities, respectively.
}
\label{fig4}
\end{figure}

We have presented SMA observations of the submillimeter 
source HDF~850.1 first discovered by Hughes et al.\ (1998)
in the HDF-N proper using SCUBA. The SMA detection yields
an extremely accurate position, which is not consistent
with previous optical/NIR identifications of the source, including 
the most recent one by Dunlop et al.\ (2004).
In fact, there is no optical or NIR counterpart
visible in the existing extremely deep imaging.
The SMA position is consistent with the positions of the 
millimeter wavelength interferometric detection of 
Downes et al.\ (1999), the weak VLA detection at 
1.4~GHz by G.~Morrison et al.\ (2009, in preparation), 
and the weak VLA detection at 8.4~GHz
by Richards et al.\ (1998).  We estimated a
millimetric redshift of $z=4.1^{+0.5}_{-0.6}$ for HDF~850.1 
by fitting the available mid-infrared through radio imaging to 
the Arp~220 SED. HDF~850.1 may be an analog of the brighter
submillimeter source GOODS~850-5 observed by Wang et al.\ (2007)
with the SMA, which also has no optical or NIR counterpart in 
extremely deep imaging (Wang et al.\ 2009) and is thought to 
lie at $z>4$ (Wang et al.\ 2007, 2009; Dannerbauer et al.\ 2008;
Daddi et al.\ 2009a).

The increasing number of such high-redshift sources now 
being identified suggests that 
they play a major role in the high-redshift star 
formation history. In Figure~\ref{fig4} we compare the 
radio powers of HDF~850.1 (green open diamond) 
and other submillimeter sources at $z>4$ (green solid symbols;
Capak et al.\ 2008 and Schinnerer et al.\ 2008; 
Coppin et al.\ 2009; Daddi et al.\ 2009a, 2009b) 
with the radio powers of a sample of spectroscopically identified 
radio-selected sources observed in several ultradeep fields 
(Cowie et al.\ 2004; Barger et al.\ 2007; 
A.~Barger et al.\ 2009, in preparation). 
We use blue open squares to denote the
optically bright radio sources ($R<24.5$) and red 
solid squares to denote the optically faint radio 
sources ($R>24.5$). While some of the high-redshift 
sources may be optically bright, many are
very faint.  The $z>4$ submillimeter sources are comparable in 
radio power to the high-luminosity end of the radio sample at $z=1-4$, 
and some lie close to the sensitivity limits of the
deepest fields observed with the VLA ($\sim20~\mu$Jy,
dot-dashed curve;
e.g., Fomalont et al.\ 2006; Owen \& Morrison 2008; 
G.~Morrison et al.\ 2009, in preparation), emphasizing 
that with the gain in 1.4~GHz sensitivity anticipated with the 
Expanded Very Large Array (EVLA; Momjian et al.\ 2009), we 
may expect such sources to be routinely included and
localized in deep 1.4~GHz samples. The EVLA will 
therefore enable the development of accurate number density and
star formation estimates from the ULIRGs at these redshifts.

\acknowledgements

We thank Glenn Morrison for permitting us to
use his VLA 1.4~GHz observation of HDF~850.1 in
advance of publication and the referee for a very thoughtful
and interesting report. We gratefully acknowledge support from 
NSF grants AST 0709356 (L.~L.~C.) and AST 0708793 (A.~J.~B.),
the Wisconsin Alumni Research Foundation and the David and 
Lucile Packard Foundation (A.~J.~B.), 
and the National Radio Astronomy Observatory (W.-H.~W.).

\end{document}